# A New and Efficient Algorithm-Based Fault Tolerance Scheme for A Million Way Parallelism

Erlin Yao, Mingyu Chen, Rui Wang, Wenli Zhang, Guangming Tan
Key Laboratory of Computer System and Architecture
Institute of Computing Technology, Chinese Academy of Sciences
Beijing, China
{yaoerlin,cmy,wangrui2009,zhangwl,tgm}@ict.ac.cn

## ABSTRACT
Fault tolerance overhead of high performance computing (HPC) applications is becoming critical to the efficient utilization of HPC systems at large scale. HPC applications typically tolerate fail-stop failures by checkpointing. Another promising method is in the algorithm level, called algorithmic recovery. These two methods can achieve high efficiency when the system scale is not very large, but will both lose their effectiveness when systems approach the scale of Exaflops, where the number of processors including in system is expected to achieve one million. This paper develops a new and efficient algorithm-based fault tolerance scheme for HPC applications. When failure occurs during the execution, we do not stop to wait for the recovery of corrupted data, but replace them with the corresponding redundant data and continue the execution. A background accelerated recovery method is also proposed to rebuild redundancy to tolerate multiple times of failures during the execution. To demonstrate the feasibility of our new scheme, we have incorporated it to the High Performance Linpack. Theoretical analysis demonstrates that our new fault tolerance scheme can still be effective even when the system scale achieves the Exaflops. Experiment using SiCortex SC5832 verifies the feasibility of the scheme, and indicates that the advantage of our scheme can be observable even in a small scale.

## 1. INTRODUCTION
Fault tolerance overhead of HPC applications is critical to the efficient utilization of contemporary HPC systems. While the peak performance of contemporary HPC systems continues to grow exponentially, it is getting more and more difficult for scientific applications to achieve high performance due to both the complex architecture of and the increasing failures in these systems [28]. Schroeder and Gibson recently studied the system logs of HPC systems in Los Alamos National Laboratory and found that the mean-time-to-interrupt (MTTI) for these systems varies from about half a month to less than half a day [19][23]. The coming Exaflops supercomputers will require the simultaneous use and control of hundreds of thousands or even millions of processing, storage, and networking elements. With this large number of elements involved, element failure will be frequent, making it increasingly difficult for applications to make forward progress. In order to use these systems efficiently and avoid restarting applications from the beginning when failure occurs, applications have to be able to tolerate failures. Fault tolerant methods attempt to enhance the utilization of system by tolerating failures occurring during the execution of applications. In this paper, we mainly discuss the fail-stop type of failures and in the processor level.

Traditional checkpointing method for the fault tolerance of HPC applications can not be scalable to a large scale. Today's long running scientific applications typically tolerate fail-stop failures by checkpointing [10][22][26]. Checkpointing can usually be used in different type of systems and to a wide range of applications. However, when applications such as High Performance Linpack (HPL) modify a large amount of memory between two consecutive checkpoints, checkpointing often introduces a considerable overhead when the number of processors used for computation is large [13]. Based on a balanced system model and statistics from the Computer Failure Data Repository (CFDR) [3], Gibson predicted that the effective application utilization of checkpoint-restart fault tolerance will keep dropping to zero under recent technology trends [24]. Today a consensus is almost reached that it will be very difficult for checkpointing to be used as the fault tolerance method of HPC systems aimed at the Exaflops.

Conventional algorithm-based fault tolerance (algorithmic recovery) method is expected to be scalable to a large scale [14], however, will still lose its efficiency when the system scale achieves the Exaflops. Compared to the checkpointing method, the most significant advantage of algorithmic recovery is that the algorithm, for the most part, runs with little modification or stoppage to perform a checkpoint, and there is no rollback of work when failure occurs. If the amount of time to recover is approximately constant relative to the overall execution time, then this greatly decreases the fault tolerance overhead. Due to these advantages, the algorithmic recovery method is expected to be scalable to a very large scale. However, as far as we know, the effectiveness of algorithmic recovery method applied to large scale has not been carefully evaluated, no matter using theoretical analysis or experiment evaluation. In this paper, based on the construction of a model for the failure and performance of system, we have investigated a theoretical analysis to the effectiveness of algorithmic recovery method applied to large scale. We found that the efficiency of algorithmic recovery method is nearly 1/3 when the system scale achieves the Exaflops.

The inefficiency of algorithmic recovery method at large scale lies in the following two reasons. First, the algorithmic recovery method is still based on the stop-and-wait scheme, where even there is only one processor failed during the execution of applications, all the processors in system have to stop and wait for recovery of the failed processor. It is clear that this stop-and-wait fault tolerance scheme can not be scalable to a very large scale. Second, with the increasing of system scale, the mean time to failure (MTTF) of system is decreasing. However, for HPC applications with intensive computation or communication, such as HPL, the cost to recover a failed processor is increasing with the system scale. When the system scale reaches a boundary, where the time to recover failure equals to or approaches the MTTF of system, it is clear that the system efficiency will be low. For algorithmic recovery method, the scale of Exaflops could be such a boundary.

To overcome these problems, this paper proposes a new algorithm-based fault tolerance scheme for HPC applications. Our major contributions are the following three points:

- First, we have established a model on the failure and performance of system. Based on this model, we have investigated a theoretical evaluation of algorithmic recovery method under large scale for the first time, and indicated the impact of the ratio between floating point computing power and communication bandwidth of processors on the efficiency of HPC system.
- Second, we have developed a non-stop scheme to tolerate fail-stop failures occurring during the execution of applications. When failure occurs during the execution, we do not stop to wait for recovery but carry out a replacement and continue the execution. At the end of execution, this scheme can recover the correct solution with a much lower cost compared to the algorithmic recovery method. To tolerate unbounded times of failures occurring during the execution, a background accelerated recovery method is proposed to rebuild the redundancy.
- Third, to demonstrate the feasibility of our new scheme to the fault tolerance for large scale HPC applications, we have incorporated our fault tolerance method to the benchmark HPL. We also have investigated a detailed efficiency analysis to this method using theoretical analysis and experiment evaluation.

Theoretical analysis demonstrates that our new fault tolerance scheme can still be effective even when the system scale achieves the Exaflops. Experiment evaluation indicates that the advantage of our scheme can be observable even in a small scale.

The rest of paper is organized as follows. Section two briefly introduces the conventional algorithm-based recovery method and demonstrates its inefficiency in the fault tolerance for Exaflops. The third part introduces our new algorithm-based fault tolerance scheme in a general framework. Section four has incorporated our scheme to the widely used benchmark HPL. In section five, a failure and performance model of system is established, based on this model, the efficiency of different fault tolerant methods under HPL has been evaluated theoretically. Section six gives some experiment results. Section seven reviews related works and provides some discussion. The last part concludes this paper and discusses some future works.

## 2. THE INEFFICIENCY OF ALGORITHM-BASED RECOVERY METHOD

In this section, first we briefly introduce the conventional algorithm-based fault tolerance scheme, then we demonstrate the inefficiency of algorithmic recovery method to tolerate fail-stop failures in large scale HPL.

### 2.1 Algorithm-Based Fault Recovery

Traditional algorithm based fault tolerance (ABFT) is proposed by Huang and Abraham, which uses information at the end of execution of an algorithm to detect and recover failures and incorrect calculations for fail-continue failures [5][21]. This technique is further developed by Chen and Dongarra to tolerate fail-stop failures occurring during the execution of HPC applications [11][12]. The idea of ABFT is to encode the original matrices using real number codes to establish a checksum type of relationship between data, and then re-design algorithms to operate on the encoded matrices to maintain the checksum relationship during the execution.

Assume there will be only one process failure, however, before the failure actually occurs, we do not know which process will fail; therefore, a scheme to handle only the lost data on the failed process actually needs to be able to handle data on any process. It seems difficult to be able to handle data on any process without saving all data on all processes somewhere. However, if we assume, at any time during the computation, the data $D_i$ on the $i$th process $P_i$ satisfies

$$D_1 + D_2 + \ldots + D_q = E \tag{1}$$

Where $q$ is the total number of processes used for computation and $E$ is data of the encoding process. Then, the lost data on any failed process would be able to be recovered from the above relationship. If the $i$th process failed during the execution, then all processes stop, and the lost data $D_i$ can be recovered from:

$$D_i = E - (D_1 + \ldots + D_{i-1} + D_{i+1} + \ldots + D_q). \tag{2}$$

In practice, this kind of special relationship is by no means natural. However, it is possible to design applications to maintain such a special checksum relationship throughout the computation, and this is one purpose of ABFT research.

### 2.2 Inefficiency of Algorithmic Recovery Method for Large Scale Systems

It can be seen from Eq. (2) that an algorithmic recovery process for a failed node needs to transfer data from other normal nodes.

Although this method avoids slow disk I/O, it depends on the network bandwidth of the HPC systems.

Based on a failure and performance model of computing system established in section 5.1, the efficiency of algorithmic recovery method for tolerating fail-stop failures in HPL is (see Section 5.2.1 for a detailed analysis):

$$E(recovery) = (1 - \frac{1}{\sqrt{p}}) e^{-0.02(8c+1)\lambda p Log(p)} . \qquad (3)$$

Where $p$ is the number of processors including in system, $\lambda$ is the failure rate of single processor, and $c$ is the ratio between floating point computing power and the network bandwidth of processors (FLOPS/bandwidth, $c$ has no unit).

A small c means a more balanced systems ($c$=1 or $c$=10, like Bluegene and Cray XT systems). For today's accelerator based HPC systems like Nebulae and Tianhe-1, the ratio $c$ may exceed 200. However a smaller $c$ may mean more total processors are needed to reach a given scale. To achieve the scale of Exaflops, a system with ratio $c$=100 may need $10^6$ processors. However, since the interconnect bandwidths of different systems are in the same magnitude, for a system with ratio $c$=10 or $c$=1, it may need $10^7$ or $10^8$ processors respectively. So in Fig. 1 we give three different upper bounds ($10^6$, $10^7$, $10^8$) of the number of processors respectively.

It can be seen that, the efficiencies of the three cases are all near to 1/3 when system achieves the scale of Exaflops, which is almost the efficiency of Triple Modular Redundancy. This indicates that the algorithmic recovery method is not suitable for the fault tolerance of HPC systems at large scale, new and efficient fault tolerance scheme should be developed for the scale of Exaflops.

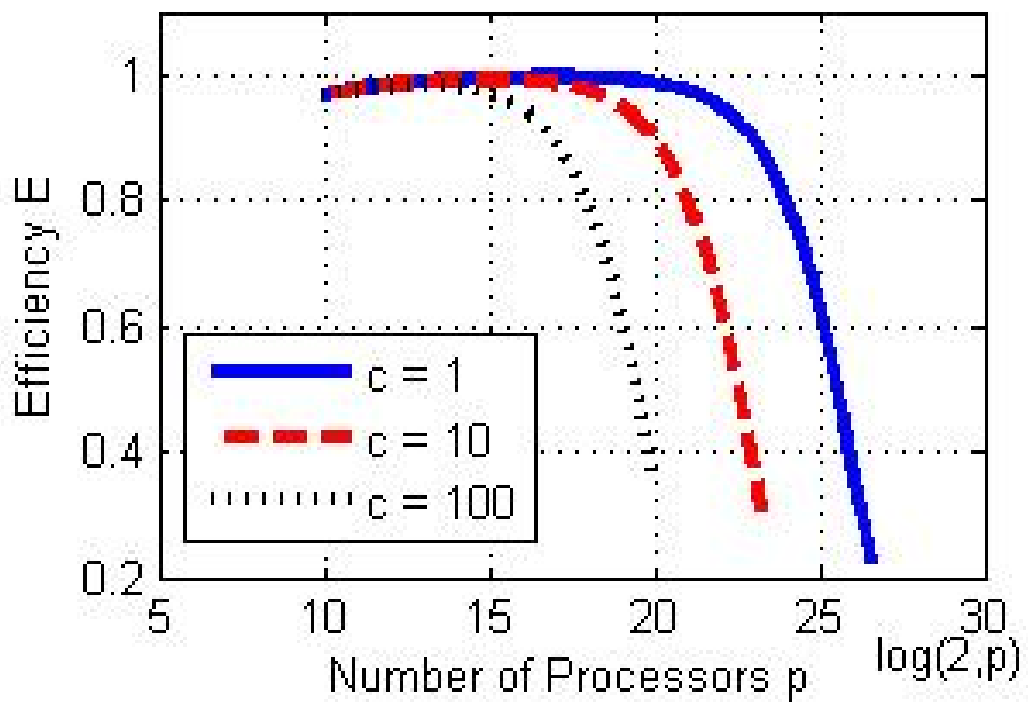

Fig. 1. Efficiency of algorithmic recovery method for fault tolerance of HPL

# 3. OUR NEW ALGORITHM-BASED FAULT TOLERANCE SCHEME

In this section, we will give some general principles for tolerating fail-stop failures in the middle of execution of applications by maintaining checksum like relationship in the algorithm instead of checkpointing or message logging.

## 3.1 Failure Detection and Location

Handling fault tolerance typically consists of three steps: 1) fault detection, 2) fault location, and 3) fault recovery or processing. Fail-stop process failures can often be detected and located with the aid of the runtime environment. For example, many current programming environments such as PVM [26], Globus [17], FT-MPI [16], and Open MPI [18] provide this type of failure detection and location capability. We assume the loss of partial processes in the message passing system does not cause the aborting of the survival processes, and it is possible to replace the failed processes in the message passing system and continue the communication after the replacement. However, the application is still responsible for recovering the data structures and the data of the failed processes. In the rest of this paper, we will mainly focus on how to process the lost data in the failed processes.

## 3.2 Failure Processing Scheme

### *3.2.1 Failure Hot-Replacement Policy*

For the simplicity of presentation, assume there will be only one process failure. Suppose at any time during the computation, the data $D_i$ on the $i$th process $P_i$ satisfies

$$D_1 + D_2 + \ldots + D_q = E.$$

Then, the lost data $D_i$ on any failed process would be able to be processed from the above relationship. Assume the $i$th process failed during the execution, instead of stopping all the processes and recovering the lost data $D_i$ using Eq. (2), we replace the failed process with the encoding process $E$ and continue the execution.

The original data is

$$D = (D_1 \ \ldots \ D_{i-1} \ D_i \ D_{i+1} \ \ldots \ D_q), \qquad (4)$$

and the transformed data (after replacement) is

$$D' = (D_1 \ \ldots \ D_{i-1} \ E \ D_{i+1} \ \ldots \ D_q). \qquad (5)$$

Then we can establish a relationship between the transformed data and the original data as:

$$D' = D \times T, \qquad (6)$$

with $T$ is a $q \times q$ matrix in the following form, where the elements omitted in the diagonal and the $i$th column are all 1, and all the other elements omitted are 0.

$$T = \begin{pmatrix} 1 & & & 1 & & & \\ & \ddots & & \vdots & & & \\ & & 1 & 1 & & & \\ & & & 1 & & & \\ & & & 1 & 1 & & \\ & & & \vdots & & \ddots & \\ & & & 1 & & & 1 \end{pmatrix} \qquad (7)$$

If the operations on data are linear transformations (such as matrix operations like decomposition), then it is clear that the relationship $D' = D \times T$ will always be kept. It can be seen that $T$ is a nonsingular matrix. At the end of execution, the original correct solution based on $D$ can be re-computed through the intermediate solution based on $D'$. It is clear that this re-computation is actually a transformation related to $T$.

However, since the redundancy decreases one after each replacement, there will be no redundancy available sooner or later if failures occurring times and times. To tolerate multiple and even unbounded times of failures during the execution, the redundancy has to be rebuilt in time. So we propose the following scheme of background accelerated recovery of redundancy.

### *3.2.2 Background Accelerated Recovery of Redundancy*

We can rebuild redundancy $E$ from

$$E = D_1 + D_2 + \ldots + D_q \quad (8)$$

It can be seen that the rebuild of redundancy is actually a recovery process, which is the same with the algorithmic recovery process. If the recovery of redundancy makes all the processors stop and wait, then it will also become the bottleneck. However, since the recovery of redundancy is not the hot spot, we can carry out recovery in the background. Then the cost on the recovery of redundancy can be overlapped with the computation or communication time during the normal execution. So the recovery process will not bring new time overhead.

But this background recovery process will bring another issue: since other processes do not stop and wait for the recovery of redundancy, when the redundancy process is rebuilt, it has fallen behind other processes in progress several steps. This may cause future stop and wait when system need synchronization. To make the whole system return to a consistent and synchronized state, some faster (such as with triple speed) processors and network need to be used to accelerate the recovery and catch up with other normal processes. Since only a small number of such high end nodes are needed, the additional cost will decrease when system scale up.

In the rest of this paper, HRBR (Hot Replace Background Recovery) will be used to represent our new fault tolerance scheme based on algorithm.

### *3.2.3 Advantages of HRBR*

Compared to the conventional failure recovery method, HRBR introduces the following two advantages. First, in a system when only one processor $P_i$ fails, all the other processors in system do not need to stop and wait for the recovery of data on $P_i$. This non-stop scheme improves the efficiency of large scale systems, since the wasted cpu cycles during stop-and-wait period increase with the scale. Second, since $T$ is a very sparse matrix, the cost of re-computing the correct solution using $T$ can be much lower than the cost of recovery of failed processor directly, especially for those computation or communication intensive applications where each processor is allocated with a large set of data. This can be seen from the example of HPL in the next section.

In the special case in section 3.2.1, we are lucky enough to be able to process the lost data on any failed process without checkpoint due to the special checksum relationship. In practice, this kind of special relationship is by no means natural. However, it is natural to ask: *is it possible to design an application to maintain such a special checksum like relationship throughout the computation on purpose?*

The following section will give some general principles.

## 3.3 A General Framework

Assume the original application is designed to run on $q$ processes. Let $D_i$ denotes the data on the $i$th computation process. In some algorithms for matrix operations (such as the Gauss Elimination algorithm for matrix decomposition), the special checksum relationship can actually be designed on purpose. The following is a general framework for fault processing:

- **Step 1:** Add another encoding process into the application. Assume the data on this encoding process is $E$. For numerical computations, $D_i$ is often an array of floating-point numbers; therefore, at the beginning of the computation, we can create a checksum relationship among the data of all processes by initializing the data $E$ on the encoding process as
  $$D_1 + D_2 + \ldots + D_q = E.$$
- **Step 2:** During the execution of the application, redesign the algorithm to operate both on the data of computation processes and on the data of encoding process in such a way that the checksum relationship is always maintained during computation.
- **Step 3:** If any process fails, then replace it with the encoding process and continue the execution.
- **Step 4:** After each replacement of failed process, rebuild the redundancy process with the background accelerated recovery method.
- **Step 5:** At the end of execution, the correct solution can be reconstructed through the intermediate result and the transformation matrix $T$.

The above fault tolerance technique can be used to tolerate fail-stop processes failures without checkpointing or message logging. The special checksum relationship between the data on different processes can be designed on purpose by using the checksum matrices of the original matrices as the input matrices. To demonstrate the feasibility of HRBR, we will apply it to the widely used benchmark HPL in the next section.

## 4. INCORPORATING HRBR INTO HPL

In this section, we apply the algorithm-based fault tolerance technique HRBR to HPL, which is one of the most important benchmarks for supercomputing and is widely used for the ranking of supercomputers on Top500 [1]. HPL is to solve system of linear equations using the Gauss Elimination with Partial Pivoting (GEPP). For the simplicity of presentation, in this section, we only discuss the case where there is only one process failed simultaneously. However, it is straightforward to extend the result here to the multiple simultaneous process failure cases by simply using a weighted checksum scheme [12].

## 4.1 The Basic Idea

We will give a simple example to demonstrate the basic idea of our method. The purpose of HPL is to solve a system of linear equations:

$$Dx=b. \quad (9)$$

During the execution, when process $P_i$ fails and after the replacement with redundancy, the linear equations become:

$$D'y=b. \quad (10)$$

At the end of execution, we get the solution $y$ as intermediate result. Since the operations in Gauss Elimination are linear transformations, it is clear that the relationship

$$D'=D\times T \quad (11)$$

is kept during the execution. Combining Eq. (9), (10) and (11), the correct solution $x$ can be calculated by:

$$x=T\times y. \tag{12}$$

If the transformation matrix $T$ is the same as (7), then expand Eq. (12), we can get:

$$\begin{cases} x_j = y_i + y_j, & if\ 1\le j\ne i\le q \\ x_i = y_i \end{cases} \tag{13}$$

It can be seen that since the simplicity of matrix $T$, the re-computing of solution $x$ is surprisingly simple.

However, in the real implementation of HPL, the encoding of data matrix and the transformation matrix $T$ will not be that simple. We will solve these issues in the next sections.

## 4.2 Two-Dimensional Block-Cyclic Data Distribution

It is well known [2] that the layout of an application's data within the hierarchical memory of a concurrent computer is critical in determining the performance and scalability of the parallel code. By using two-dimensional block-cyclic data distribution [2], HPL seeks to maintain load balance and reduce the frequency with which data must be transferred between processes.

For reasons described above, HPL organizes the one-dimensional process array representation of an abstract parallel computer into a two-dimensional rectangular process grid. Therefore, a process in HPL can be referenced by its row and column coordinates within the grid. An example of such an organization is shown in Fig. 2.

| | | | | | |
|---|---|---|---|---|---|
| 0 | 1 | 2 | 3 | 4 | 5 |

(a)

| | 0 | 1 | 2 |
|---|---|---|---|
| 0 | 0 | 1 | 2 |
| 1 | 3 | 4 | 5 |

(b)

Fig. 2. Process grid in HPL. (a) One-dimensional process array. (b) Two-dimensional process grid.

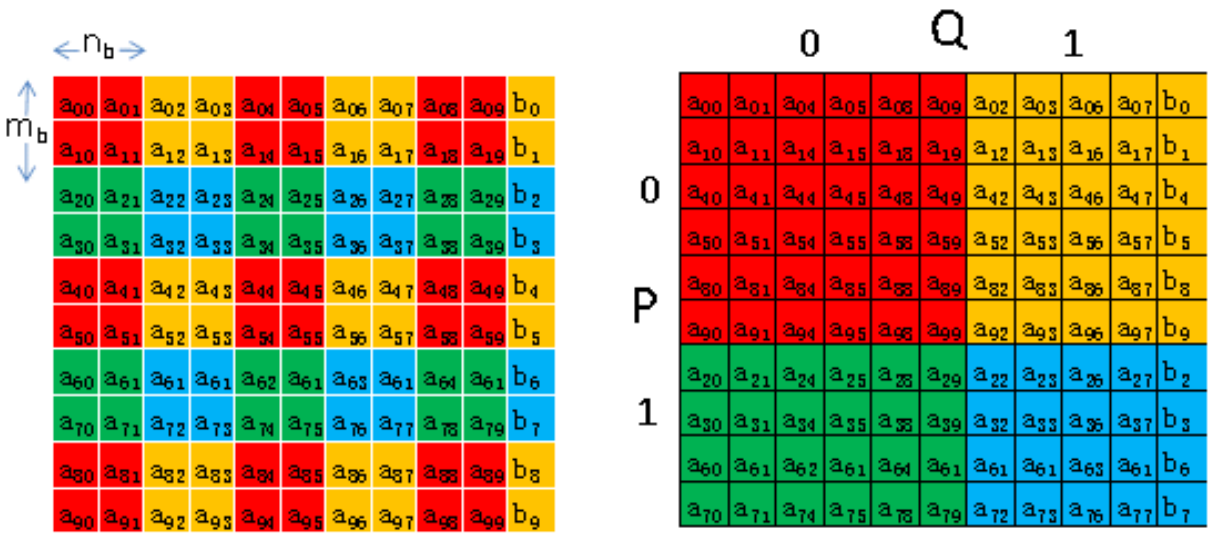

Global View  Local (distributed) View

Fig. 3. Two-dimensional block-cyclic matrix distribution.

The two-dimensional block-cyclic data distribution scheme is a mapping of the global matrix onto the rectangular process grid. There are two pairs of parameters associated with the mapping. The first pair of parameters is ($m_b$, $n_b$), where $m_b$ is the row block size and $n_b$ is the column block size. The second pair of parameters is ($P$, $Q$), where $P$ is the number of process rows in the process grid and $Q$ is the number of process columns in the process grid.

Given an element $a_{ij}$ in the global matrix $A$, the process coordinate ($p_i$, $q_j$) that $a_{ij}$ resides can be calculated by

$$\begin{cases} p_i = \lfloor i/m_b \rfloor \ \bmod\ P, \\ q_j = \lfloor j/n_b \rfloor \ \bmod\ Q. \end{cases} \tag{14}$$

The local coordinate ($i_{pi}$, $j_{qj}$) which $a_{ij}$ resides in the process ($p_i$, $q_j$) can be calculated according to the following formula:

$$\begin{cases} i_{p_i} = \lfloor \lfloor i/m_b \rfloor / P \rfloor * m_b + (i \ \bmod\ m_b), \\ j_{q_j} = \lfloor \lfloor j/n_b \rfloor / Q \rfloor * n_b + (j \ \bmod\ n_b). \end{cases} \tag{15}$$

For HPL, in the mapping of data onto the process grid, the right hand side $b$ is viewed as a column vector besides the coefficients matrix $A$: ($A|b$). Fig. 3 is an example of mapping a 10×10 matrix and $b$ onto a 2×2 process grid according to two-dimensional block-cyclic data distribution with $m_b$=$n_b$=2.

## 4.3 Encoding Data Distribution Matrix

In this section, we will construct encoding scheme which can be used to incorporate the HRBR method into the fault tolerance of HPL. The purpose of encoding is to create the checksum relationship proposed in Step 1 in Section 3.3.

Suppose the dimension of coefficients matrix $A$ is $n$, $n$ has achieved the magnitude of $2\times10^6$ for today's fastest supercomputer [1]. Since $b$ is very small relative to $A$, we treat the redundancy of $A$ and $b$ separately and we mainly discuss the redundancy scheme of $A$ in this paper. The redundancy of $b$ can be simply viewed as there are multiple copies of $b$.

| | | | |
|---|---|---|---|
| 0 | 1 | 0 | 1 |
| 2 | 3 | 2 | 3 |
| 0 | 1 | 0 | 1 |
| 2 | 3 | 2 | 3 |

(a)

| | | | |
|---|---|---|---|
| 0 | 0 | 1 | 1 |
| 0 | 0 | 1 | 1 |
| 2 | 2 | 3 | 3 |
| 2 | 2 | 3 | 3 |

(b)

Fig. 4. Two-dimensional block-cyclic distribution of an example matrix. (a) Original matrix from global view. (b) Original matrix from distributed view.

| | | | | | |
|---|---|---|---|---|---|
| 0 | 1 | 1 | 0 | 1 | 1 |
| 2 | 3 | 5 | 2 | 3 | 5 |
| 0 | 1 | 1 | 0 | 1 | 1 |
| 2 | 3 | 5 | 2 | 3 | 5 |

(a)

| | | | | | |
|---|---|---|---|---|---|
| 0 | 0 | 1 | 1 | 1 | 1 |
| 0 | 0 | 1 | 1 | 1 | 1 |
| 2 | 2 | 3 | 3 | 5 | 5 |
| 2 | 2 | 3 | 3 | 5 | 5 |

(b)

Fig. 5. Distributed row checksum matrix of the original matrix. (a) Row checksum matrix from global view. (b) Row checksum matrix from distributed view.

| | | | |
|---|---|---|---|
| 0 | 1 | 0 | 1 |
| 2 | 3 | 2 | 3 |
| 0 | 1 | 0 | 1 |
| 2 | 3 | 2 | 3 |

(a)

| | |
|---|---|
| 1 | 1 |
| 5 | 5 |
| 1 | 1 |
| 5 | 5 |

(b)

| | |
|---|---|
| 1 | 0 |
| 1 | 0 |
| 0 | 1 |
| 0 | 1 |

(c)

Fig. 6. Global view of data and redundancy matrix. (a) Global view of matrix $A$. (b) Global view of redundancy $A*V$. (c) Coding matrix $V$.

Assume the data matrix $A$ is originally mapped onto a $P\times Q$ process grid according to the two-dimensional block-cyclic data distribution. We adopt the same encoding scheme introduced in [12] which is developed for the conventional algorithmic recovery method. For the convenience of presentation, assume the size of the local matrices in each process is the same. We will explain our coding scheme for the matrix $A$ with the help of the example matrix in Fig. 4. Fig. 4a shows the global view of all the elements of the example matrix. After the matrix is mapped onto a 2×2 process grid with $m_b$=$n_b$=1, the distributed view of this matrix is shown in Fig. 4b.

Suppose we want to tolerate a single process failure. We dedicate another $P$ additional processes and organize the total ($PQ+P$) processes as a $P\times(Q+1)$ process grid with the original matrix $A$ distributed onto the first $P$ rows and $Q$ columns of the process grid.

The distributed row checksum matrix $A^r$ of the matrix $A$ is the original matrix $A$ plus the part of data on the ($Q$+1)th process columns which can be obtained by adding all local matrices on the first $Q$ process columns. Fig. 5b shows the distributed view of the row checksum matrix of the example matrix from Fig. 4. Fig. 5a is the global view of the row checksum matrix.

Generally the redundancy (in columns) of matrix $A$ can be viewed as $A*V$, where $V$ is often called the coding matrix. For the simple case of row checksum, $V$ is an $n\times1$ vector with

$$V= (1\ 1\ \ldots\ 1)^T \qquad (16)$$

Here $V^T$ denotes the transpose of matrix $V$. For the example matrix $A$ in Fig. 4, the global view of redundancy $A*V$ and the coding matrix $V$ are as Fig. 6 shown.

## 4.4 Maintaining Encoding Relationship during the Execution

In this section, we will show that the encoding relationship between the original data matrix and the redundancy matrix established at the initial step can be maintained during the execution of HPL.

Solving system of linear equations using the Gauss Elimination method can also be viewed as a process of LU decomposition of the coefficients matrix. Given a matrix $A$, the purpose of LU decomposition is to find a lower triangular matrix $L$ and an upper triangular matrix $U$ such that $A$=$L*U$. Then the system of equations $Ax$=$b$ is equivalent to solving $Ux$=$L^{-1}b$. Since $L$ and $U$ are both triangular matrix, then $Ax$=$b$ can be solved by simple back substitution using $L$ and $U$ with a low cost.

The LU decomposition is a recursive process. The starting point is $L_0$=$I_n$ (the identity matrix) and $U_0$=$A$. At the end of decomposition, we get the lower triangular matrix $L_n$=$L$, and the upper triangular matrix $U_n$=$U$. At each of the intermediate step ($L_s$, $U_s$), does the encoding relationship between the data matrix and redundancy still be maintained? Then the following conclusion can be given:

**Theorem 1.** *In HPL, assume ($A|b|AV$) is the original data matrix and redundancy matrix. If at each step of the LU decomposition, we let the redundancy matrix executes the same update operations as the U matrix, then after each step ($L_s$, $U_s$), the redundancy matrix is $U_s*V$, i.e., the encoding relationship holds between the U matrix and redundancy.*

**Proof.** If we view $L_0$=$I_n$ and $U_0$=$A$, then the initial step is

$$(A|b|AV) = (U_0\ |\ L_0^{-1}*b|\ U_0*V). \qquad (17)$$

After the first step of decomposition, $L_0$ becomes $L_1$ with

$$L_1= L_0* l_1, \qquad (18)$$

where $l_1$ is a matrix and the update operations on matrix $M$ using $l_1$ is equivalent to $M$ left multiplied by matrix $l_1^{-1}$: $l_1^{-1}*M$. Since $U_0$, $L_0^{-1}*b$ and $U_0*V$ execute the same update operations using $l_1$, so $U_0$ becomes $U_1$ with

$$U_1= l_1^{-1}*U_0; \qquad (19)$$

$L_0^{-1}*b$ becomes: $l_1^{-1}*L_0^{-1}*b = L_1^{-1}*b$;

and $U_0*V$ becomes

$$l_1^{-1}*U_0*V = U_1*V. \qquad (20)$$

So the encoding relationship is kept between the redundancy matrix and the $U$ matrix. And it is clear that the solving of $Ax$=$b$ is equivalent to the solving of $U_1 x = L_1^{-1} b$.

Similarly, we can prove that the encoding relationship holds after each of the step ($L_s$, $U_s$) with $0\le s\le n$. □

## 4.5 Failure Replacement Policy and Transformation Matrix

### 4.5.1 *Failure Replacement Policy*

When some process fails, assume its position in the process grid is ($p$, $q$), instead of only replacing it with the corresponding redundant process in the $p$th row, we replace all the $p$ processes in the $q$th column with the entire column of redundant processes.

The reason that we adopt this replacement policy is as follows: if we replace the entire $q$th column of processes with the column of redundant processes, then in the global view of matrix $A$, the replacement can be viewed as some columns of $A$ are replaced with the redundant columns $A*V$, and this type of replacement can be easily represented as a linear transformation of $A$. However, if we only replace the failed process with some redundant process, then the replacement can not be represented as a transformation of matrix $A$, so will bring difficulty to the deriving of transformation matrix, which can be seen in the next section.

### 4.5.2 *The Deriving of Transformation Matrix*

In general, we can assume that the redundancy of matrix $A$ is $A*V$, where $V$ is an $n\times m$ matrix, and can be viewed as $m$ columns of redundancy vectors. As discussed in the above section, the replacement after failure can be viewed as the interchange of some columns of $A$ with the $m$ columns of redundancy vectors $A*V$. After the replacement, matrix $A$ becomes $A'$, as discussed above, there should exist a transformation matrix $T$ such that $A' =A*T$. Then what is the form of matrix $T$? Lemma 1 is clear.

According to Lemma 1, it can be seen that the $i_1$, $i_2$,…, $i_m$ columns of $T$ are exactly the encoding vectors $V_1$, $V_2$,…, $V_m$ respectively. The elements omitted in the diagonal of $T$ (Eq. (22)) are all 1.

As for the example matrix $A$ in Fig. 6, if process(0,1) fails, we replace process(0,1) and process(1,1) with the corresponding redundant processes, then the matrix $A'$ after replacement and the transformation matrix $T$ are as Fig. 7 shown.

According to $x$=$T*y$ and the form of $T$ in Fig. 7, the re-computing of $x$ can be expanded as the following:

$$x_1=y_1+y_2,\ x_2=y_2,\ x_3=y_3+y_4,\ x_4=y_4. \qquad (21)$$

**Lemma 1.** *For an n× n matrix A, suppose the redundant columns are an n × m matrix: $(A*V_1|A*V_2|\ldots|A*V_m)$. If the $i_1$, $i_2,\ldots,i_m$ columns of A are interchanged with these redundant columns respectively, then A becomes a matrix A′. There exists a matrix T such that A′=A*T, where T is an n× n matrix in the following form:*

$$T = \begin{pmatrix} 1 & & & V_{1,1} & & & & V_{m,1} & & \\ & \ddots & & \vdots & & & & V_{m,2} & & \\ & & 1 & V_{1,i_1-1} & & & & \vdots & & \\ & & & V_{1,i_1} & & & & & & \\ & & & V_{1,i_1+1} & 1 & & & \vdots & & \\ & & & \vdots & & \ddots & & & & \\ & & & & & & \ddots & \vdots & & \\ & & & \vdots & & & 1 & V_{m,i_m-1} & & \\ & & & & & & & V_{m,i_m} & & \\ & & & \vdots & & & & V_{m,i_m+1} & 1 & \\ & & & V_{1,n-1} & & & & \vdots & & \ddots \\ & & & V_{1,n} & & & & V_{m,n} & & 1 \end{pmatrix} \quad (22)$$

(a)

| | | | |
|---|---|---|---|
| 0 | 1 | 0 | 1 |
| 2 | 3 | 2 | 3 |
| 0 | 1 | 0 | 1 |
| 2 | 3 | 2 | 3 |

(b)

| | | | |
|---|---|---|---|
| 0 | 1 | 0 | 1 |
| 2 | 5 | 2 | 5 |
| 0 | 1 | 0 | 1 |
| 2 | 5 | 2 | 5 |

(c)

| | | | |
|---|---|---|---|
| 1 | 1 | 0 | 0 |
| 0 | 1 | 0 | 0 |
| 0 | 0 | 1 | 1 |
| 0 | 0 | 0 | 1 |

Fig. 7. Global view of replacement and transformation. (a) Global view of matrix *A*. (b) Global view of matrix *A*′. (c) Transformation matrix *T*.

# 5. THEORETICAL ANALYSIS

In this section, we will establish a model for the failure and performance of system, based on which we will theoretically analyze the efficiency of different methods for the fault tolerance of HPL, including algorithmic recovery method and HRBR.

## 5.1 Failure and Performance Model of System

### *5.1.1 Model of System Failure*

In this paper we mainly discuss the type of fail-stop failures, and in the processor level. Suppose one computing system consists of *p* processors, we assume that all the processors are homogeneous, so have the same MTTF (assume as *M*). Then the failure rate λ of single processor is: $\lambda=1/M$. To achieve the scale of Exaflops, the number of processors included in a system is expected to be more than one million. In this paper we make an optimistic assumption that single processor's MTTF is 10 years, which is about $3.15\times10^8$ seconds. If a system consists of $10^6$ processors, then MTTF of the system will be: $3.15\times10^8/10^6=315$ seconds. This means that there will be a failure occurring about five minutes. In contrast, many high performance scientific applications require running of hours and even days.

Suppose *t* is the time of failure occurs of one processor and $D(t)$ is the failure probability distribution function for single processor, since the exponential distribution is the only continuous random distribution with a constant failure rate [6], we assume that $D(t)$ is an exponential distribution, so

$$D(t)=1-e^{-\lambda t}. \quad (23)$$

The probability that there is no failure of one processor during time interval [0, *t*] is: $1-D(t)=e^{-\lambda t}$. We make a simple assumption that the failures of processors in a system are independent from each other, so the probability that all the *p* processors in system have not failed during time interval [0, *t*] is:

$$(1-D(t))^p=e^{-\lambda pt}. \quad (24)$$

The system is viewed as failed if there is one or more processors failed, so the probability that the system is failed during time interval [0, *t*] is exactly:

$$1-(1-D(t))^p=1-e^{-\lambda pt}. \quad (25)$$

### *5.1.2 Performance Model of System Running HPL*

For each processor in system, we assume that its floating point computing power is *f* (flops), and its memory size is *m* (byte). Then the total amount of memory of system is *mp*, and the theoretical peak performance of system is *fp*. According to a balanced system model and the long standing trends in top500's statistics of supercomputer [1], it is reasonable to assume that:

$$mp : fp = 0.4 \quad (26)$$

We use the IEEE 754 double precision standard of floating point number, so the size of every number is 8 bytes. Suppose *n* is the matrix dimension in HPL, then the size of coefficients matrix is $8n^2$ bytes, and the whole computation amount is about $2n^3/3$ times of floating point operations. According to the rule of thumb for choosing the matrix dimension [2], 80% of the total amount of memory should be filled by the coefficients matrix, so it holds that:

$$8n^2=0.8pm \quad (27)$$

If all the processors in system are fully utilized and there is no failure occurs during the execution, the whole run time of HPL should be about:

$$T= 2n^3 / (3fp) \quad (28)$$

However, due to the dropping of system's MTTF, often failures will occur during the execution of applications. Fault tolerant method is adopted to tolerate failures during the execution, and this will result in extra cost in the execution time of applications and overhead in the system hardware. Suppose the execution of HPL will last $T'$ ($T' \geq T$) time under some fault tolerant method and the hardware efficiency of system under this method is *e*, then we can define the efficiency of this fault tolerant method as:

$$E(\text{method})= e\times T/T' \quad (29)$$

## 5.2 Efficiency of Algorithmic Recovery

As discussed above, the main overhead of algorithmic recovery method is the recovery of data on the failed processor. As for HPL, the coefficients matrix is evenly distributed onto all the processors. Suppose the size of coefficients matrix is $n^2$, then the amount of data distributed onto one single processor is $n^2/p$ floating point numbers. As described in section 4.2, the *p* processors in system are organized as a $P\times Q$ process grid. In general, *P* and *Q* are slightly different. In the convenience of description, we assume that $P$=$Q$, so $P$=$Q$ =$p^{0.5}$. The recovery of any corrupted number involves the participation of $Q$ processors using

$$D_i = E-(D_1 +\ldots+ D_{i-1} + D_{i+1}+\ldots+ D_Q). \qquad (30)$$

Since these $Q$ processors can compute and communicate in parallel, it is clear that one recovery requires Log($Q$) times of floating point operations and communications. In section 5.1.2, we have assumed that the floating point computing power of single processor is $f$, if the ratio of $f$ and the communication bandwidth (bytes/second) between processors is $c$, then the recovery of one floating point number (8 bytes) is

$$\frac{Log(Q)}{f}+\frac{8cLog(Q)}{f}=\frac{(8c+1)Log(p)}{2f} \qquad (31)$$

So the cost of time to recover all the data on one failed processor is

$$\frac{n^2}{p}\times\frac{(8c+1)Log(p)}{2f}=\frac{(8c+1)n^2Log(p)}{2pf}. \qquad (32)$$

However, failures can also occur during the period of recovery. Since the states of processors used for recovery have not changed during the recovery, we can restart recovery from the beginning when failure occurs during the recovery. Then in expectation how long will the recovery cost? According to the property of exponential distribution [6], if we let $t$= $(8c+1)n^2$Log($p$)/(2$pf$), then in expectation the cost of recovery of one failed processor is

$$t'=\frac{e^{\lambda pt}-1}{\lambda p}=\frac{e^{\lambda p\times\frac{(8c+1)n^2Log(p)}{2pf}}-1}{\lambda p}=\frac{e^{\frac{(8c+1)\lambda n^2Log(p)}{2f}}-1}{\lambda p} \qquad (33)$$

Since the recovery process only begins when there is a failure occurring, and the mean time between failures is $M/p$, so the time efficiency of algorithmic recovery method is

$$\frac{M/p}{M/p+t'}. \qquad (34)$$

According to the encoding scheme of algorithmic recovery, it is clear that its hardware efficiency is

$$e=1-\frac{1}{Q}=1-\frac{1}{\sqrt{p}}. \qquad (35)$$

According to the definition on the efficiency of fault tolerant method (Eq. (29)), the efficiency of algorithmic recovery is

$$E(recovery)=e\times\frac{M/p}{M/p+t'}=(1-\frac{1}{\sqrt{p}})e^{\frac{-(8c+1)\lambda n^2Log(p)}{2f}} \qquad (36)$$

According to Eq. (26) and Eq. (27), it holds that $n^2$=0.04$pf$, replacing into Eq. (36), we have

$$E(recovery)=(1-\frac{1}{\sqrt{p}})e^{-0.02(8c+1)\lambda pLog(p)} \qquad (37)$$

The relationship between the efficiency of algorithmic recovery method and the system scale is as Fig. 1 shown.

## 5.3 Efficiency Analysis of HRBR

### *5.3.1 Overhead of Background Accelerated Recovery*

As discussed in section 3.2.2, we can reserve some faster processors and network to help the background recovery and speeding of redundancy processor. Then what is the overhead of this background accelerated recovery method? First we need the following lemma:

**Lemma 2.** *Suppose a system consists of p homogeneous processors, the failure rate of each processor is λ, and the failures of different processors are independent from each other. During a time interval [0, t], the probability that there are more than one processor failed is*

$$1+(p-1)e^{-\lambda pt}-pe^{-\lambda t(p-1)} \qquad (38)$$

**Proof.** It is clear that during time interval [0, $t$], the probability that there are exactly $k$ ($0\le k\le p$) processors failed is

$$P_k=\binom{p}{k}(1-e^{-\lambda t})^k(e^{-\lambda t})^{p-k}. \qquad (39)$$

And it holds that

$$\sum_{k=0}^{p}P_k=1. \qquad (40)$$

So it is clear that $P_0$=$e^{-\lambda pt}$, $P_1$=$p(1-e^{-\lambda t})e^{-\lambda t(p-1)}$, and the probability that there are more than one processor failed is $1-P_0-P_1$. □

Then the following theorem can be given:

**Theorem 2.** *For a computing system running HPL, suppose the system consists of p processors, the ratio of computing power and communication bandwidth between processors is c, and system's failure rate is λp. Assume*

$$s=e^{0.02(8c+1)\lambda pLog(p)} \qquad (41)$$

*If the same redundancy method as in section* 4 *is adopted and initially we build three columns of redundancy, if the background accelerated recovery adopt faster processors and network with speed of s times of the original speed, then the execution of HPL can be completed with a probability of about* 0.98.

**Proof.** Under the speed of original processor and network, the recovery time of one redundancy is $t'$ (from Eq. (33)). In the background accelerated recovery, assume the speed of faster processors and network is $s$ times of the original speed, then the recovery time of one redundancy is $t'/s$. During the recovery of redundancy, the processors in progress have not stopped to wait. So to catch up with these processors, it will cost the redundancy processor another time of $t'/(s(s-1))$. So in fact the rebuild of one redundancy will cost time

$$\frac{t'}{s}+\frac{t'}{s(s-1)}=\frac{t'}{(s-1)}. \qquad (42)$$

If we let this rebuild time $t'/(s-1)$=$M/p$, according to Eq.(33), it can be seen that

$$s=e^{0.02(8c+1)\lambda pLog(p)} \qquad (43)$$

In the background accelerated recovery, if we adopt faster processors and network with $s$ times of the original speed, then in expectation the rebuild time of one redundancy is equal to the MTTF of system. This means that in expectation, during the period of one processor failed, another processor can be repaired, which constitutes a balance of system.

Since there are three columns of redundancy built initially, then the system is failed only when there are more than three processors failed during one rebuild period *M/p*, according to Lemma 2, this probability is

$$1-P_0-P_1-P_2-P_3 \quad (44)$$

with $t = M/p$, and $P_k = \binom{p}{k}(1-e^{-\lambda t})^k (e^{-\lambda t})^{p-k}$. (from Eq. (39))

Then it can be calculated that the probability is about 0.02, so the execution of HPL can be completed with a probability of 0.98. □

It can be seen that the required speedup *s* is increasing with the number of processors *p*. When $p=10^6$ and $c=100$, $s\approx 2.7$, so we let $s=3$ is enough. Note that the analysis of rebuilding process here is not in detail due to its complexity, but the overhead of redundancy has already been considered.

### *5.3.2 Efficiency of HRBR*

As discussed in section 4.1, the re-computing of correct solution involves only the following computation: $x_j=y_i+y_j$ or $x_i=y_i$. So the re-computing only processes *n* times (recovering solution *x*) of floating point operations. If we assume that the floating point computing power of single processor is *f*, and the ratio of *f* and the network bandwidth (bytes/second) between processors is *c*, then the re-computing of one floating point number (8 bytes) is

$$\frac{1}{f}+\frac{8\times c}{f}=\frac{8c+1}{f}. \quad (45)$$

So the time cost to re-compute all the *n* floating point numbers is

$$t=(8c+1)n/f. \quad (46)$$

Since the re-computing process only begins when there is a failure occurring, and the mean time between failures is *M/p*, so the time efficiency of HRBR is

$$\frac{M/p}{M/p+t}=\frac{M}{M+0.04(8c+1)p^2/n} \quad (47)$$

According to the analysis in section 5.3.1, there will be six columns of processors for redundancy and background accelerated recovery, then the hardware efficiency of HRBR is

$$e=1-\frac{6}{Q}=1-\frac{6}{\sqrt{p}} \quad (48)$$

According to the definition on the efficiency of fault tolerant method (Eq. (29)), the efficiency of HRBR should be

$$E(HRBR)=0.98(1-\frac{6}{\sqrt{p}})\frac{M}{M+0.04(8c+1)p^2/n} \quad (49)$$

In Eq. (49), if we let $c=100$ and consider three cases: $n=p$, $n=10p$ and $n=100p$, then we can get the efficiency of HRBR as Fig. 8 shown. According to Fig. 8, in the worst case ($n=p$) the efficiency of HRBR is still above 0.88 when the system scale reaches one million processors, while then the efficiency of algorithmic recovery is nearly 1/3.

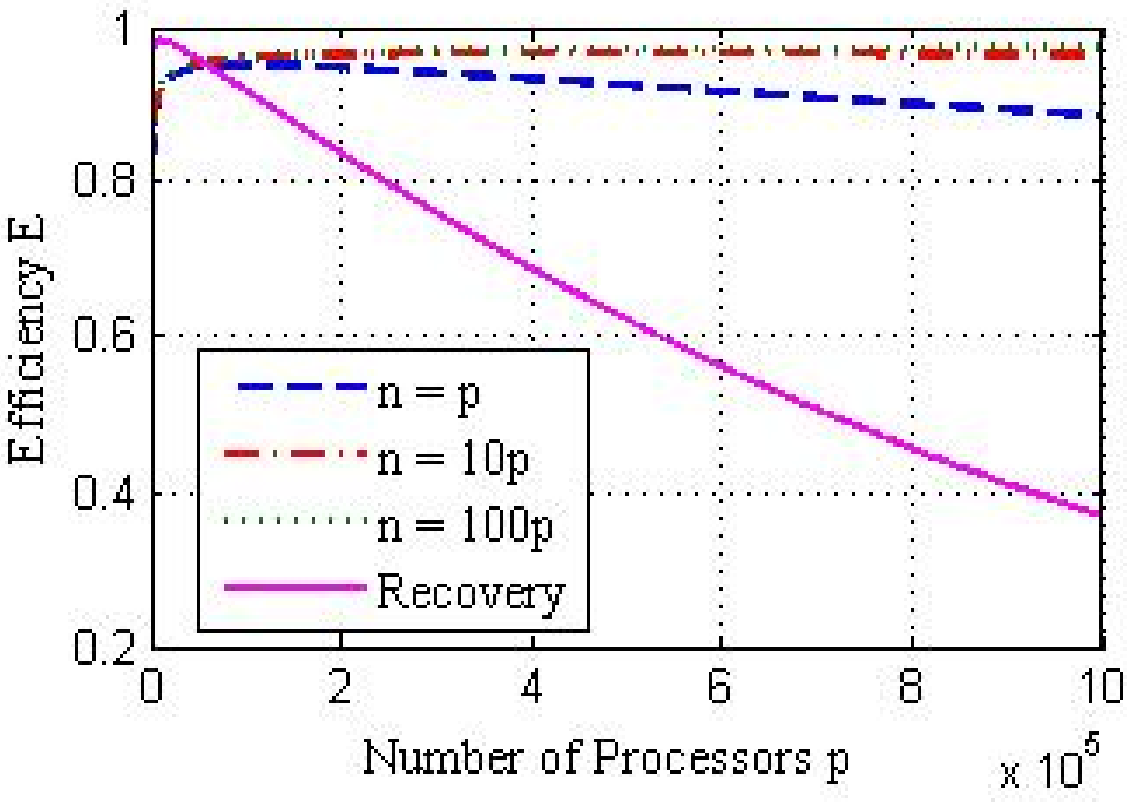

Fig. 8. Efficiency of HRBR and algorithmic recovery method for fault tolerance of HPL.

## 6. EXPERIMENT EVALUATION

We perform a set of experiments to verify the feasibility of HRBR method and compare its performance with the algorithm-based recovery method.

The experiments were performed on a SiCortex SC5832 supercomputer. The SC5832 has 972 nodes, 5,832 cores and 3888 GB of memory. It uses a diameter-6 Kautz graph for 2,916 links. Each node consists of six cores that implement the MIPS64 instruction set architecture (ISA). The cluster interconnect provides a maximum bandwidth of 2 GB/s. The reason that we choose SC5832 as the platform lies in that: first, it's a very balanced system (the ratio between computing power and communication bandwidth is nearly 2); second, its network is very stable, making the fluctuations in experiments nearly negligible.

**Table 1:** Comparison between the performance of HRBR and algorithmic recovery method for 3-times recovery

| *N* | *P* | HRBR total time (s) | HRBR GFlops | ABFT-R total time (s) | ABFT-R GFlops |
|---|---|---|---|---|---|
| 147000 | 48*48 | 2873.05 | 7.372e+02 | 2931.62 | 7.224e+02 |
| 112000 | 36*36 | 2202.62 | 4.253e+02 | 2236.79 | 4.188e+02 |
| 93000 | 30*30 | 1775.22 | 3.021e+02 | 1825.17 | 2.939e+02 |
| 74000 | 24*24 | 1439.57 | 1.877e+02 | 1474.89 | 1.832e+02 |
| 55000 | 18*18 | 1048.86 | 1.058e+02 | 1066.13 | 1.040e+02 |
| 36000 | 12*12 | 710.15 | 4.381e+01 | 718.89 | 4.327e+01 |

Our experiments are not based on a full implementation since it needs support from runtime system for failure detection and recovery. In [27] we have discussed the needed MPI feature for self–healing MPI programming.

To make things simple, we use scheduled failures during the HPL update phases. Failures were simulated by flushing the matrix data and marking the status of processes to be dead. All failures could be detected by a reduce operation on the values of process status after every update phase. And a new failure is only generated after failure handlings of the previous one has been finished. Both algorithmic recovery and HRBR method are implemented with MPICH2.

To implement background accelerated recovery of redundancy, we need some faster processors and network. We use openMP to help "build" faster processors. We initiate one process per physical core and leave a free core for each redundant process. When it needs to accelerate computation, the redundant process run in OpenMP mode with two thread, so that it seems as if it was running on a twice faster processor. As for the faster network, we add additional accelerated rows of processes in the process grid to help compute processes reduce their local matrices to redundant processes.

In the experiments we keep the local matrix size to about 3000x3000 and change the system scale up to 2300 cores. Three successive failures are scheduled in each test. The Recovery and HRBR mode are compared in the same configuration except some additional cores for HRBR acceleration.

Experiment results show that both methods get the correct results with negligible error. Table 1 shows the comparison between the performance of HRBR and algorithmic recovery method, where $N$ is the matrix dimension and $P$ is the number of processes totally used. It can be seen that under the scale of experiment ($P$ ranges from the order of $10^2$ to $10^3$), the gap between the performance of HRBR and algorithmic recovery method is small, however, the advantage of HRBR is still observable.

# 7. RELATED WORKS AND DISCUSSION

## 7.1 Related Works

Conventional algorithm based fault tolerance (ABFT) uses information at the end of execution of an algorithm to detect and recover failures and incorrect calculations for fail-continue failures [5] [21]. Fail-continue failures are failures which do not halt execution. The idea of ABFT is to encode the original matrices using real number codes and then re-design algorithms to operate on the encoded matrices. In [21], Huang and Abraham proved that the encoding relationship in the input encoded matrices is preserved at the end of the computation no matter which algorithm is used to perform the computation. Therefore, processor miscalculations can be detected, located, and corrected at the end of the computation. ABFT algorithms have received wide investigation, including development for specific hardware architectures [8], efficient codes for use in ABFT [7] [25], and analysis of potential performance under more advanced failure conditions [4].

In a distributed environment such as high performance clusters, if a failed processor stops working, then we need to be able to tolerate failures in the middle of the computation, to recover the data or adopt other possible methods. Recently, the use of algorithmic methods without checkpointing to help recover during fail-stop failures has been investigated by Chen and Dongrra [11][12]. It has been shown that this method can be applied to large scale parallel matrix operations, like matrix multiplication, Cholesky factorization, and iterative methods for solving linear equations[13][14][15]. By incorporating this method into ScaLAPACK and HPL, experimentation demonstrates that this method has decreasing overhead in relation to overall runtime as the matrix size increases, and thus shows promise to reduce the expected runtime for HPC applications on very large matrices.

## 7.2 Discussion

In this paper, a new algorithm-based fault tolerant scheme for fail-stop failures has been proposed, which has combined the advantages of Huang's ABFT method [21] for fail-continue failures and Chen's algorithmic recovery method [12] for fail-stop failures. The recovery of correct solution is performed at the end of the execution, and the redundancy is recovered in background immediately when failure and replacement occur. Though in this paper we only demonstrate how to incorporate HRBR to the HPL, it is clear that HRBR can be easily extended to matrix operations such as addition, multiplication, scalar product and transposition. The advantage of HRBR lies in that it has dramatically decreased the impact of failures on the execution of applications with hot replacement. Note that if an algorithmic recovery process has no hotspot, the background recovery mechanism can also be used.

The failure modeling in this paper is not new since a lot of studies have done in this field. However, the main purpose of failure modeling is to compare the efficiency of different methods. We only discuss the single processor failure, where the failure mode like a mid-plane or rack failure has not been considered. But this kind of failures can be tolerated by carefully designing the redundancy pattern. In fact, our scheme can endure multiple nodes failure if they are not in the same row. Future work will extend the failure mode to more practical issues.

Some high end system may have a longer node-MTTF than in this paper, and the failures rates of future systems may not be as high as estimated. However, if only the premise that "MTTF of system is decreasing with the increasing of system scale" holds, the proposed HRBR will stand on its own when the system scale is large enough. High end system may only postpone but not solve the scale problem.

# 8. CONCLUSION AND FUTURE WORKS

In this paper, we have presented a new algorithm-based fault tolerance scheme, hot replacement with background accelerated recovery, for HPC applications. Because no periodical checkpoint or stop-and-wait recovery is involved in this approach, process failures can often be tolerated with a very low overhead. We showed the practicality of this technique by applying it to the HPL, which is representative benchmark for supercomputers to achieve high performance and scalability. Theoretical analysis and experiment evaluation demonstrate that our new scheme can still be efficient for a million way parallelism at the Exaflops scale, when traditional checkpointing and even algorithmic recovery method failed.

There are many directions in which this work could be extended. The first direction is to extend this approach to more applications. Second, a more robust implementation of this scheme at large scale and under real circumstances should be developed. Furthermore, it is also interesting to investigate the accuracy and stability of this algorithm-based fault tolerance technique at large scale.

## 9. ACKNOWLEDGMENTS
This research was supported in part by the National Natural Science Foundation of China (61003062). We appreciate the Argonne National Laboratory for the use of SiCortex SC5832.